\documentclass{iopjournal}

\usepackage{amsmath}
\usepackage{amssymb}
\usepackage{graphicx}
\usepackage{booktabs}
\usepackage{xcolor}
\usepackage[nameinlink,capitalise]{cleveref}

\graphicspath{{./}}
\crefname{figure}{Fig.}{Figs.}
\Crefname{figure}{Fig.}{Figs.}
\crefname{table}{Table}{Tables}
\Crefname{table}{Table}{Tables}

\begin{document}

\articletype{Paper}

\title{Multi-objective Bayesian optimisation of a double-layer target for quasi-monoenergetic TNSA protons}

\author{Cheng-Qi Zhang$^1$, Yang He$^2$, Mamat Ali Bake$^2$ and Bai-Song Xie$^{1,*}$}

\affil{$^1$Key Laboratory of Beam Technology of the Ministry of Education, and School of Physics and Astronomy, Beijing Normal University, Beijing 100875, China}

\affil{$^2$Xinjiang Key Laboratory of Solid State Physics and Devices, School of Physics Science and Technology, Xinjiang University, Urumqi 830017, China}

\email{bsxie@bnu.edu.cn}

\begin{abstract}
We carry out a six-parameter multi-objective Bayesian optimisation of a carbon--hydrogen double-layer target for target-normal-sheath proton acceleration. The campaign consists of 80 two-dimensional EPOCH simulations with the laser amplitude $a_0$, pulse duration $\tau$, carbon-layer thickness $L_1$, hydrogen-layer density $N_2$, hydrogen-layer thickness $L_2$ and hydrogen-layer radius $r_p$ as input variables. Each final proton spectrum is scored by the peak energy, the charge fraction inside a $\pm10\%$ peak-energy window and the charge in that window. Among the Pareto-set evaluations, the cases with peak energies between 64 and 71 MeV occur near $a_0=30$, $\tau=45$ fs, $L_1=0.3\,\mu{\rm m}$, $L_2=30$ nm and $r_p=0.15\,\mu{\rm m}$. Along this branch, increasing $N_2$ raises the in-window charge and increases the bandwidth. The small rear-layer radius keeps the proton source within the flat central region of the transverse sheath field, where the accelerating field is nearly uniform. A 3D calculation is performed for the intermediate-density case $N_2=11.85\,n_c$, which balances bandwidth and in-window charge along this branch. The corresponding 2D spectrum has $E_{\rm peak}=67.4$ MeV and $\Delta E/E=18.8\%$, whereas the 3D spectrum has $E_{\rm peak}=34.1$ MeV and $\Delta E/E=7.0\%$. The lower 3D peak energy and narrower bandwidth are associated with an earlier decay of the rear-sheath field and an earlier saturation of the proton peak energy, and the quasi-monoenergetic peak is retained in 3D.
\end{abstract}

\section{Introduction}

Laser-plasma ion acceleration is a widely studied route to compact sources of short proton and ion bunches~\cite{snavely2000intense,hatchett2000electron}. Laser-accelerated protons have been used or proposed for proton radiography and field probing~\cite{borghesi2002electric}, isochoric heating of dense matter~\cite{patel2003isochoric}, fast-ignition concepts~\cite{roth2001fast}, radiotherapy-related concepts~\cite{bulanov2002feasibility,linz2007will}, and radiobiology or FLASH-motivated studies~\cite{zeil2013dose,lv2022flash,kroll2022tumour}. In these settings, the energy spectrum matters. A narrow energy peak gives a better-defined stopping range in matter~\cite{iaea2005radiation,linz2007will}, helps spectral and spatial beamline selection~\cite{brack2020spectral}, and places more source charge in the energy interval selected for an experiment. Quasi-monoenergetic proton beams with useful charge are therefore a common objective in laser-ion acceleration studies~\cite{hegelich2002mev,schwoerer2006microstructured,haberberger2012collisionless}.

Laser-driven ion acceleration has been investigated in several regimes~\cite{daido2012review,macchi2013ion}, which include radiation-pressure acceleration~\cite{esirkepov2004highly,macchi2013ion}, collisionless-shock acceleration~\cite{fiuza2012laser,haberberger2012collisionless}, breakout-afterburner acceleration~\cite{yin2007monoenergetic}, and magnetic-vortex acceleration~\cite{nakamura2010high}. Target-normal-sheath acceleration (TNSA) has been studied extensively and is the mechanism considered here~\cite{wilks2001energetic,mora2003plasma,passoni2010target}. In TNSA, the laser heats electrons at the front of a solid target. A fraction of these electrons crosses the foil and forms a charge-separation sheath that accelerates ions from the rear surface~\cite{wilks2001energetic,mora2003plasma}. However, the sheath field is nonuniform in space and time, so flat foils usually produce broad proton spectra.

Several target designs have been studied to control the spectral shape, and progress in sub-micrometre three-dimensional lithography and multi-material nanoprinting is making increasingly small and precise structures realisable~\cite{skliutas2025multiphoton,yang2021multimaterial,wen2021silica}. Scaling studies have examined the dependence on laser parameters and target thickness~\cite{mora2005thin,fuchs2006scaling,robson2007scaling}, and experiments have shown the role of preplasma conditions~\cite{wagner2016maximum,higginson2018near}and prepulse control~\cite{kaluza2004influence}. Microstructured targets can localise the proton source~\cite{hegelich2002mev,schwoerer2006microstructured}, near-critical-density structures change the electron and field distribution near the target~\cite{sgattoni2012laser}, and nanostructured layers can modify absorption and sheath formation~\cite{passoni2016nanostructured,bin2015ion}. Double-layer targets were introduced to separate the sheath-forming heavy-ion substrate from the light-ion population intended for acceleration~\cite{esirkepov2002double}. Here the heavy substrate produces the hot electrons and the rear sheath, while the small light-ion layer occupies only the central part of that sheath. Because this central region is more uniform, the protons gain similar energies and form a narrower spectrum. This sets up a charge--bandwidth trade-off: a smaller or thinner source narrows the spectrum but carries little charge, whereas a denser source carries more charge but broadens the peak through beam-loading and space-charge effects~\cite{neely2006enhanced,mckenna2008effects}.

Balancing these quantities requires a search over laser and target parameters, and each kinetic simulation is expensive. Bayesian optimisation (BO) is suited to this type of problem because it builds a probabilistic surrogate model and uses an acquisition function to choose new evaluations~\cite{jones1998efficient,rasmussen2006gaussian,snoek2012practical,shahriari2016taking,srinivas2010gaussian}. In laser-plasma acceleration, BO has been applied to laser-wakefield accelerators~\cite{shalloo2020automation,jalas2021bayesian}, to numerical optimisation of laser-driven ion acceleration~\cite{smith2020optimizing,dolier2022multi}, to experimental closed-loop control of laser-driven proton beams~\cite{loughran2023automated}, and to TNSA pulse-shaping calculations~\cite{zhang2026doublepulse}. When several beam quantities are relevant, multi-objective BO keeps the nondominated set available for analysis after the campaign~\cite{irshad2023multi,irshad2024pareto}. These studies have optimised the maximum or peak proton energy; here the spectral bandwidth and the charge within a peak-energy window are retained as explicit objectives alongside the peak energy.

In this work, we use multi-objective BO to optimise a carbon--hydrogen double-layer target for quasi-monoenergetic TNSA protons. The optimisation varies six laser and target parameters, including the rear-layer density, thickness and transverse radius, which directly affect the charge--purity trade-off discussed above. Each 2D PIC evaluation is assessed by the peak energy, the charge fraction inside a $\pm10\%$ peak-energy window and the charge in that window, which keeps energy, spectral concentration and charge as separate objectives. The completed data set contains 80 2D evaluations. A 3D calculation is then performed for one point on the high-energy branch to examine how the spectrum and sheath evolution change.

\section{Simulation and optimisation method}\label{sec:methods}

\subsection{Simulation setup}\label{sec:setup}

The numerical experiments are performed with the relativistic particle-in-cell code EPOCH~\cite{arber2015epoch}. The target consists of an exponential front preplasma, a fully ionized carbon layer and a rear-side hydrogen layer with finite transverse radius. The fixed preplasma scale length is $L_g=0.5\,\mu{\rm m}$, and the carbon electron density is $N_1=50\,n_c$, where $n_c$ is the critical density for the laser wavelength $\lambda_0=0.8\,\mu{\rm m}$. The laser is linearly polarized and has a Gaussian transverse profile with fixed spot size $w_0=3\,\mu{\rm m}$.

The free input vector is
\begin{equation}
  \mathbf{x}=(a_0,\tau,L_1,N_2,L_2,r_p),
\end{equation}
where $a_0$ is the normalized laser amplitude, $\tau$ is the full-width at half-maximum pulse duration, $L_1$ is the carbon thickness, $N_2$ and $L_2$ are the density and thickness of the rear hydrogen layer, and $r_p$ is its transverse radius. A summary of the input ranges is given in \cref{tab:variables}. The upper bounds $a_0=30$ and $\tau=45$ fs correspond to a pulse energy of about 13 J for the fixed spot size, within reach of current high-power laser systems~\cite{danson2019petawatt}.

\begin{table*}[!ht]
  \centering
  \caption{Laser and target input parameters and their search ranges.}
  \label{tab:variables}
  \begin{tabular}{lll}
    \toprule
    Parameter & Meaning & Search range \\
    \midrule
    $a_0$ & Normalized laser amplitude & $5$--$30$ \\
    $\tau$ & Pulse duration & $25$--$45$ fs \\
    $L_1$ & Carbon-substrate thickness & $0.3$--$1.0\,\mu{\rm m}$ \\
    $N_2$ & Hydrogen-layer density & $0.5$--$30\,n_c$ \\
    $L_2$ & Hydrogen-layer thickness & $30$--$200$ nm \\
    $r_p$ & Hydrogen-layer radius & $0.15$--$1.5\,\mu{\rm m}$ \\
    \bottomrule
  \end{tabular}
\end{table*}

The 2D optimisation simulations use a box $x\in[0,100]\,\mu{\rm m}$ and $y\in[-10,10]\,\mu{\rm m}$ with $n_x=10000$ and $n_y=1000$, corresponding to $\Delta x=10$ nm and $\Delta y=20$ nm. Each cell is initialised with 20 electron, 10 carbon and 500 proton macroparticles. The laser enters from the left boundary and open boundaries are used for particles and fields. All 2D evaluations are stopped at $t=400$ fs, by which the proton peak energy has nearly saturated, and use the same resolution and box. 

\subsection{Bayesian optimisation}\label{sec:bo}

Each final proton spectrum is reduced to three objective quantities. The spectrum is weighted by the macroparticle charge and binned in energy. The peak energy $E_{\rm peak}$ is the dominant high-energy feature of this distribution. A peak window $0.9E_{\rm peak}\le E \le 1.1E_{\rm peak}$ defines the in-window charge $Q_{\rm in}$, and $Q_{\rm tot}$ is the charge of the forward-going protons ($p_x>0$) above 5 MeV. The three objectives are
\begin{equation}
  O_1=E_{\rm peak},\qquad
  O_2=Q_{\rm in}/Q_{\rm tot},\qquad
  O_3=\log_{10}Q_{\rm in}.
  \label{eq:objectives}
\end{equation}
All three objectives are maximised: $O_2$ is the in-window purity, the fraction of the analysed charge inside the peak window, and $O_3$ is the in-window charge, which distinguishes spectra of similar purity.

The bandwidth of the peak is reported as a separate diagnostic:
\begin{equation}
  \Delta E/E={\rm FWHM}/E_{\rm peak}
\end{equation}
where the full width at half maximum (FWHM) is the energy width of the peak, measured between the two points at which the spectrum falls to half of its maximum value. 

Bayesian optimisation fits a probabilistic surrogate to the evaluated data and uses an acquisition function to decide where to sample next~\cite{shahriari2016taking}. Each objective is modelled independently with a Gaussian process using a Mat\'ern-$5/2$ covariance kernel with separate length scales for each input, after the inputs are normalised to the unit cube and the objective values are standardised~\cite{rasmussen2006gaussian}. The kernel hyperparameters are obtained by maximising the marginal likelihood.

\begin{figure*}[!ht]
  \centering
  \includegraphics[width=\textwidth]{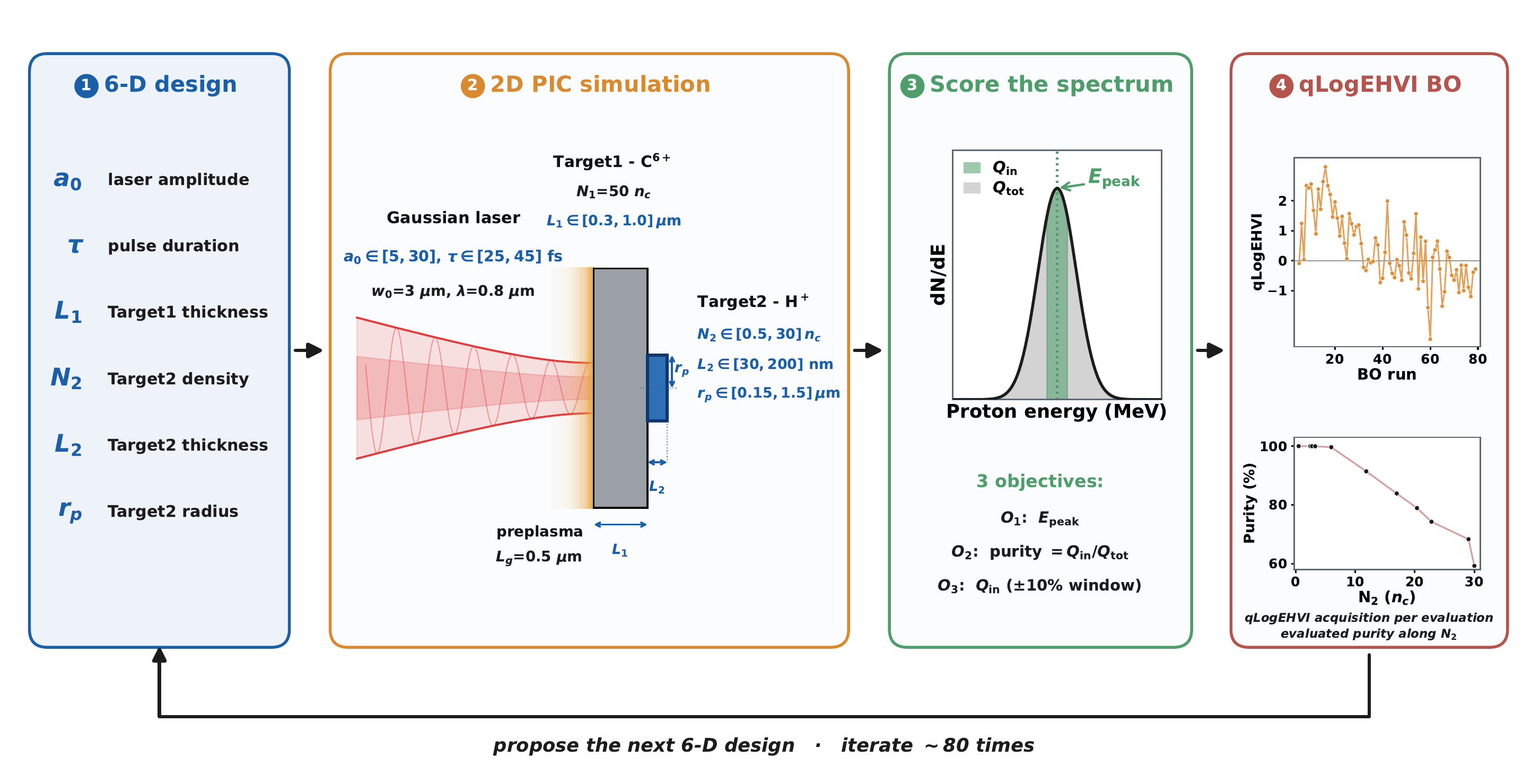}
  \caption{Multi-objective Bayesian-optimisation workflow for the double-layer target, shown as four repeated steps. Step 1: the six-dimensional design vector. Step 2: each parameter set is evaluated with a 2D PIC simulation of a C$^{6+}$ substrate and a finite-radius rear H$^+$ layer, sketched with its fixed quantities. Step 3: the proton spectrum is scored by three maximised objectives, the peak energy $E_{\rm peak}$, the in-window purity $Q_{\rm in}/Q_{\rm tot}$, and the in-window charge $Q_{\rm in}$ in the $\pm10\%$ peak-energy window. Step 4: the qLogEHVI acquisition function proposes the next parameter set from the Gaussian-process surrogate, and the loop repeats for about 80 evaluations.}
  \label{fig:workflow}
\end{figure*}
For several competing objectives, the aim is to enlarge the set of nondominated solutions, measured by the hypervolume that the objective values dominate with respect to a fixed reference point~\cite{irshad2023multi}. New parameter sets are chosen by maximising the expected hypervolume improvement, the gain in dominated hypervolume that a candidate parameter set is predicted to add. This acquisition is computed with qLogEHVI~\cite{daulton2020differentiable,daulton2021parallel,ament2023unexpected}, a parallel, differentiable and numerically stable logarithmic form of that improvement, as implemented in BoTorch~\cite{balandat2020botorch}. The reference point is $(0,0,10)$ in the objective coordinates $(E_{\rm peak},Q_{\rm in}/Q_{\rm tot},\log_{10}Q_{\rm in})$, a lower corner at zero peak energy, zero purity and $\log_{10}Q_{\rm in}=10$, chosen low enough that every successful evaluation lies above it and contributes to the volume. Failed evaluations are mapped below this corner and add no hypervolume, but are retained as records rather than discarded. The campaign is initialised with space-filling samples and continued with acquisition-selected batches, giving 80 2D PIC evaluations in total; the complete loop of parameter-set proposal, 2D PIC evaluation, spectral scoring and acquisition-driven update is summarised in \cref{fig:workflow}. In the last step of that figure, the upper plot shows the qLogEHVI value of the parameter set chosen at each of the 80 evaluations, where larger values mark the more promising candidates. The lower plot shows the scored purity of the evaluated parameter sets against $N_2$, a one-dimensional projection of the six-parameter search shown for $N_2$ because the high-purity solutions vary mainly along it. The Pareto set is then computed from the completed records, and the fitted Gaussian processes are also sampled to map the dependence of each objective on the inputs.

\section{Results and discussion}\label{sec:results}

\begin{figure*}[!ht]
  \centering
  \includegraphics[width=\textwidth]{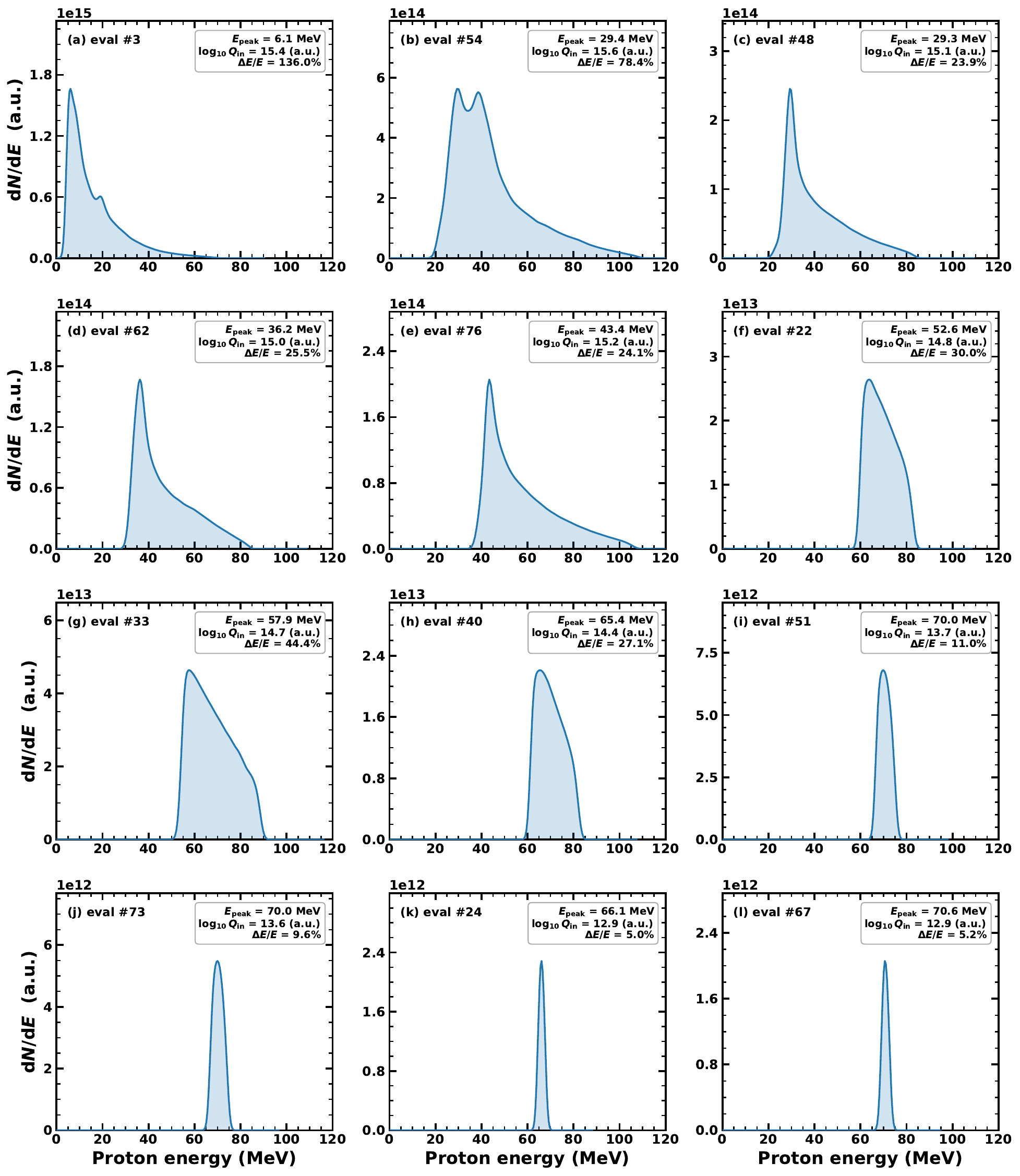}
  \caption{Representative proton spectra from the 6D multi-objective Bayesian-optimisation campaign. Twelve completed evaluations are selected from the full record after sorting by increasing spectral purity.}
  \label{fig:spectra}
\end{figure*}

The optimiser samples spectra ranging from broad, multi-feature distributions to narrow single peaks. Twelve completed evaluations, ordered by increasing spectral purity, are collected in \cref{fig:spectra}. Low-purity cases are broad or multi-peaked, whereas high-purity cases show a distinct peak near 66--71 MeV and occur for small rear-layer radii and thin hydrogen layers.

\begin{figure}[!ht]
  \centering
  \includegraphics[width=0.6\columnwidth]{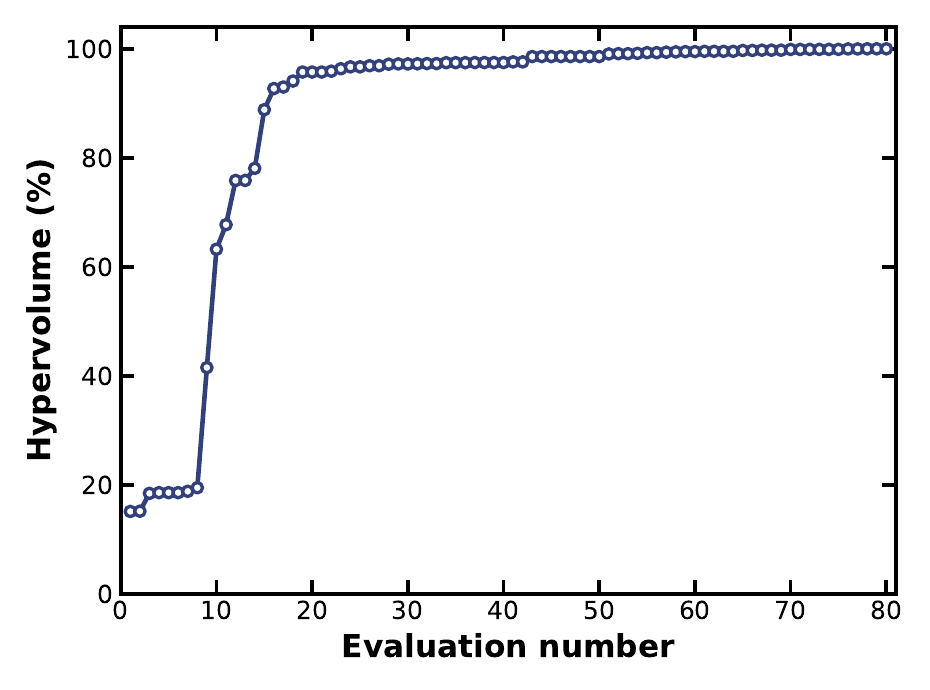}
  \caption{Convergence of the optimisation. The cumulative dominated hypervolume is plotted as a percentage of its final value over the 80 PIC evaluations. The three objectives are $E_{\rm peak}$, purity and $\log_{10}Q_{\rm in}$.}
  \label{fig:hv}
\end{figure}

The dominated hypervolume summarises the whole nondominated set with a single number. In the three-objective space it is the volume that the nondominated points dominate above the reference point, and it grows as the front reaches higher peak energy, purity and in-window charge and spreads further over the trade-off. It converges within the early part of the campaign, rising steeply over the initial evaluations and reaching a plateau after about 30, as \cref{fig:hv} shows. The completed data set contains 34 Pareto-optimal evaluations. The later runs mainly refine and populate the trade-off surface.

\begin{figure*}[!ht]
  \centering
  \includegraphics[width=\textwidth]{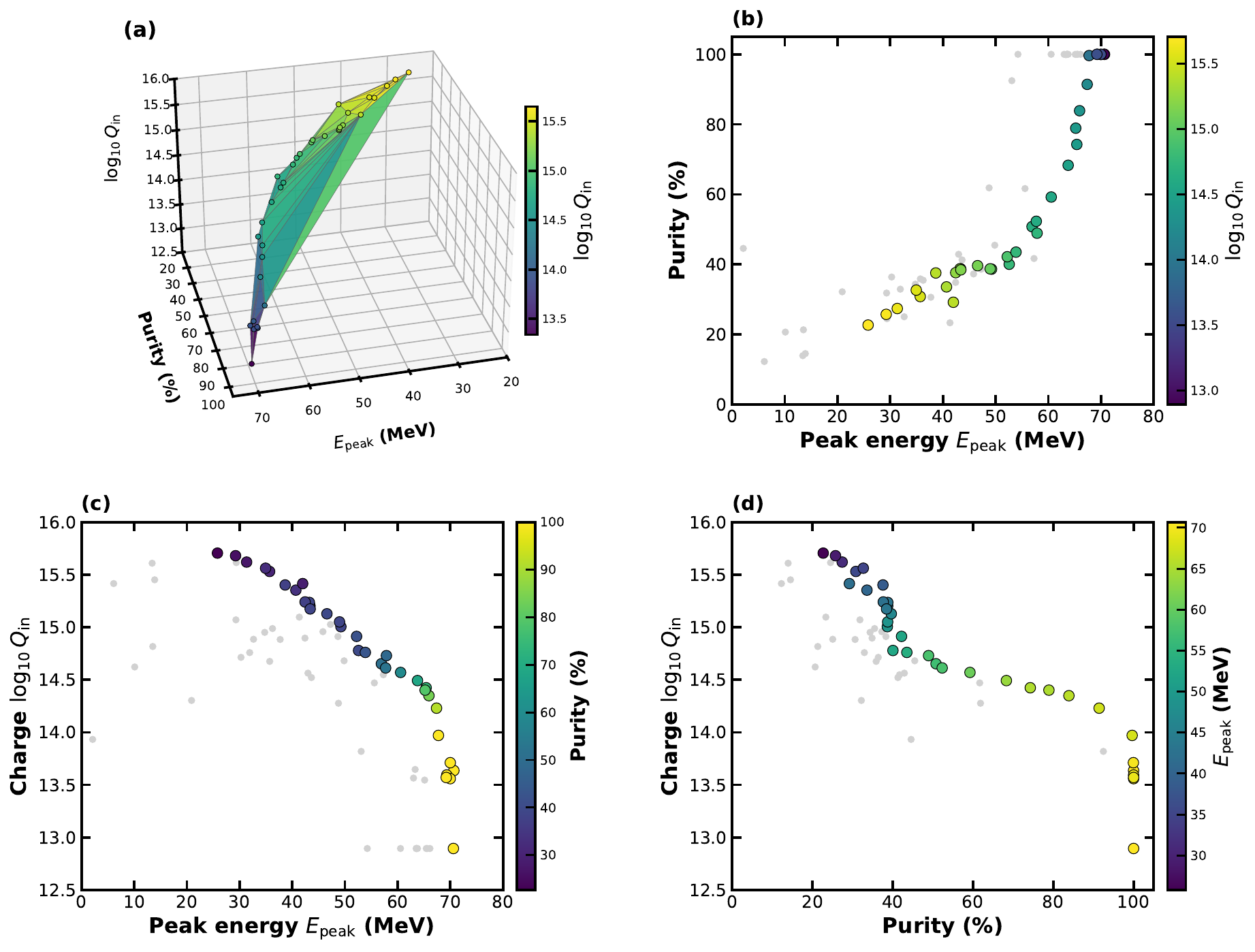}
  \caption{Three-objective Pareto front and two-dimensional projections. (a) Pareto-optimal evaluations in the objective space of peak energy $E_{\rm peak}$, spectral purity and peak-window charge $\log_{10}Q_{\rm in}$. Panels (b)--(d) show the corresponding two-dimensional projections. Grey points denote dominated evaluations and coloured points denote Pareto-optimal parameter sets; the colour scale encodes the third objective not shown on the two axes. Evaluations \#42, \#50 and \#16 are labelled along the $N_2$ branch.}
  \label{fig:pareto}
\end{figure*}

The nondominated solutions, displayed in the three-dimensional objective space and in three two-dimensional projections in \cref{fig:pareto}, separate high-purity, low-charge spectra from broader spectra with larger in-window charge. Keeping the three objectives separate preserves this charge--purity variation across neighbouring parameter sets.

In the high-energy region of the Pareto set, the points are concentrated at the bounds $a_0=30$, $\tau=45$ fs, $L_1=0.3\,\mu{\rm m}$, $L_2=30$ nm and $r_p=0.15\,\mu{\rm m}$. The remaining variation is mainly along the hydrogen-layer density $N_2$. Higher purity favours the smallest source, pushing $L_1$, $L_2$ and $r_p$ to their lower limits, while a higher peak energy and purity favour a hotter electron population, driving $a_0$ and $\tau$ to their upper limits; no energy budget is imposed, and the pulse energy at this corner, about 13 J, is simply the value set by those bounds. This leaves $N_2$ as the only parameter free to trade charge against bandwidth along the front. The Gaussian-process main effects vary monotonically up to these bounds and show no interior optimum. The labelled points \#42, \#50 and \#16 are three cases on this $N_2$ branch. Evaluation 50 is also used for the 3D calculation because it lies between the lower-density cases with smaller peak-window charge and the higher-density cases with broader bandwidth.

\begin{figure*}[!ht]
  \centering
  \includegraphics[width=0.61\textwidth]{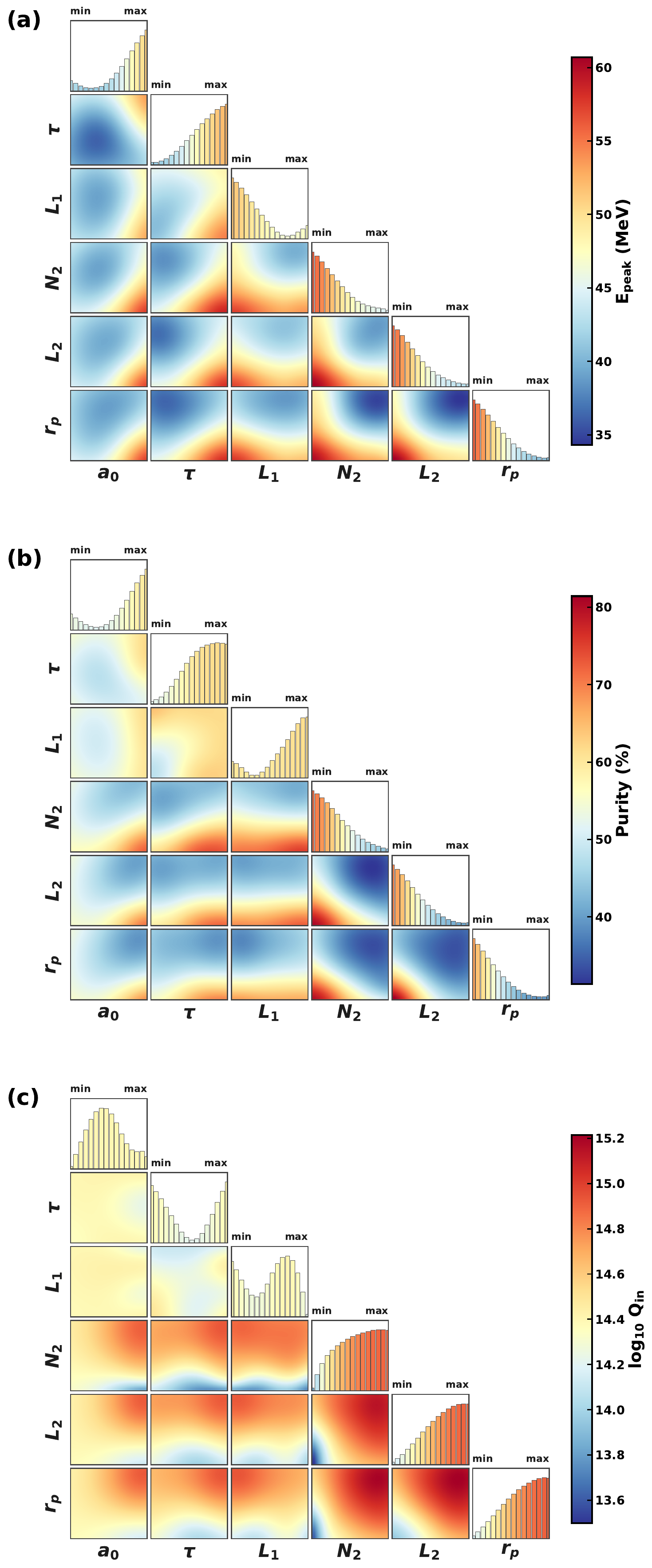}
  \caption{Gaussian-process response maps over the six-dimensional design space. Three surrogate-model corner plots show the predicted dependence of (a) peak energy $E_{\rm peak}$, (b) spectral purity and (c) $\log_{10}Q_{\rm in}$ on the six inputs $a_0$, $\tau$, $L_1$, $N_2$, $L_2$ and $r_p$. For each objective, a Mat\'ern-$5/2$ Gaussian process is fitted to the 80 completed evaluations. Diagonal panels show one-dimensional main effects, while lower-triangle panels show pairwise posterior-mean maps with the other inputs marginalised over the evaluated parameter sets.}
  \label{fig:input}
\end{figure*}

The fitted surrogate maps the dependence of $E_{\rm peak}$, purity and $\log_{10}Q_{\rm in}$ on the six inputs, displayed in \cref{fig:input}. The trends below are read from the posterior means within the sampled domain. The two laser variables act mainly on the peak energy. A higher $a_0$ and a longer $\tau$ both raise $E_{\rm peak}$, consistent with a larger absorbed energy and a hotter electron population driving a stronger rear sheath; the same direction is associated with higher purity, as a stronger and more sharply peaked sheath produces a more pronounced spectral peak.

Among the target variables, a thinner carbon substrate $L_1$ increases both $E_{\rm peak}$ and purity. A shorter electron transport length to the rear surface reduces the transverse spreading and cooling of the hot electrons before they form the sheath, giving a stronger and more localised accelerating field. A thinner hydrogen layer $L_2$ also favours higher purity, because a thinner proton sheet starts from a narrower range of initial positions in the sheath and therefore acquires a narrower final energy spread.

The hydrogen-layer density $N_2$ and radius $r_p$ control the spectral shape most directly. Increasing $N_2$ raises $\log_{10}Q_{\rm in}$ but lowers purity, as a denser proton population carries more charge but loads the sheath and broadens the peak through space-charge and beam-loading effects. A smaller $r_p$ raises purity with little effect on $E_{\rm peak}$, since a narrower source samples only the central, more uniform part of the rear sheath. Together these dependences place the high-energy, high-purity solutions at small $L_1$, $L_2$ and $r_p$, with $N_2$ as the remaining free parameter.

Once the high-energy Pareto points have reached the bounds listed above, $N_2$ acts as the main tuning parameter along the branch. \Cref{tab:ridge} summarises selected points on this branch. Because the 2D geometry is invariant along the out-of-plane direction, the in-window charge is a line density and is reported per micron of out-of-plane length. At low density, the rear hydrogen layer behaves closer to a test-particle source and gives high purity with little charge. Evaluation 20 has $\Delta E/E=5.24\%$ and purity 1.000, but only $1.26$ pC/$\mu{\rm m}$ in the peak window. As $N_2$ increases, $Q_{\rm in}$ rises while the spectrum broadens. Between evaluations 42 and 16, the peak energy changes from 67.74 to 65.94 MeV, while the bandwidth increases from 14.11\% to 22.96\% and the in-window charge increases from 14.97 to 35.76 pC/$\mu{\rm m}$. This part of the Pareto set is mainly a charge--bandwidth variation at nearly fixed peak energy.

\begin{table*}[!ht]
  \centering
  \caption{Selected 2D Pareto points on the high-energy $N_2$ branch.}
  \label{tab:ridge}
  \begin{tabular}{lcccccc}
    \toprule
    Eval. & $N_2$ ($n_c$) & $E_{\rm peak}$ (MeV) & FWHM (MeV) & $\Delta E/E$ (\%) & Purity & $Q_{\rm in}$ (pC/$\mu$m) \\
    \midrule
    \#20 & 0.50  & 70.58 & 3.70  & 5.24  & 1.000 & 1.26 \\
    \#42 & 5.98  & 67.74 & 9.56  & 14.11 & 0.996 & 14.97 \\
    \#50 & 11.85 & 67.38 & 12.70 & 18.84 & 0.914 & 27.21 \\
    \#16 & 16.97 & 65.94 & 15.14 & 22.96 & 0.839 & 35.76 \\
    \#40 & 22.80 & 65.41 & 17.74 & 27.12 & 0.743 & 42.53 \\
    \bottomrule
  \end{tabular}
\end{table*}

\begin{figure*}[!ht]
  \centering
  \includegraphics[width=\textwidth]{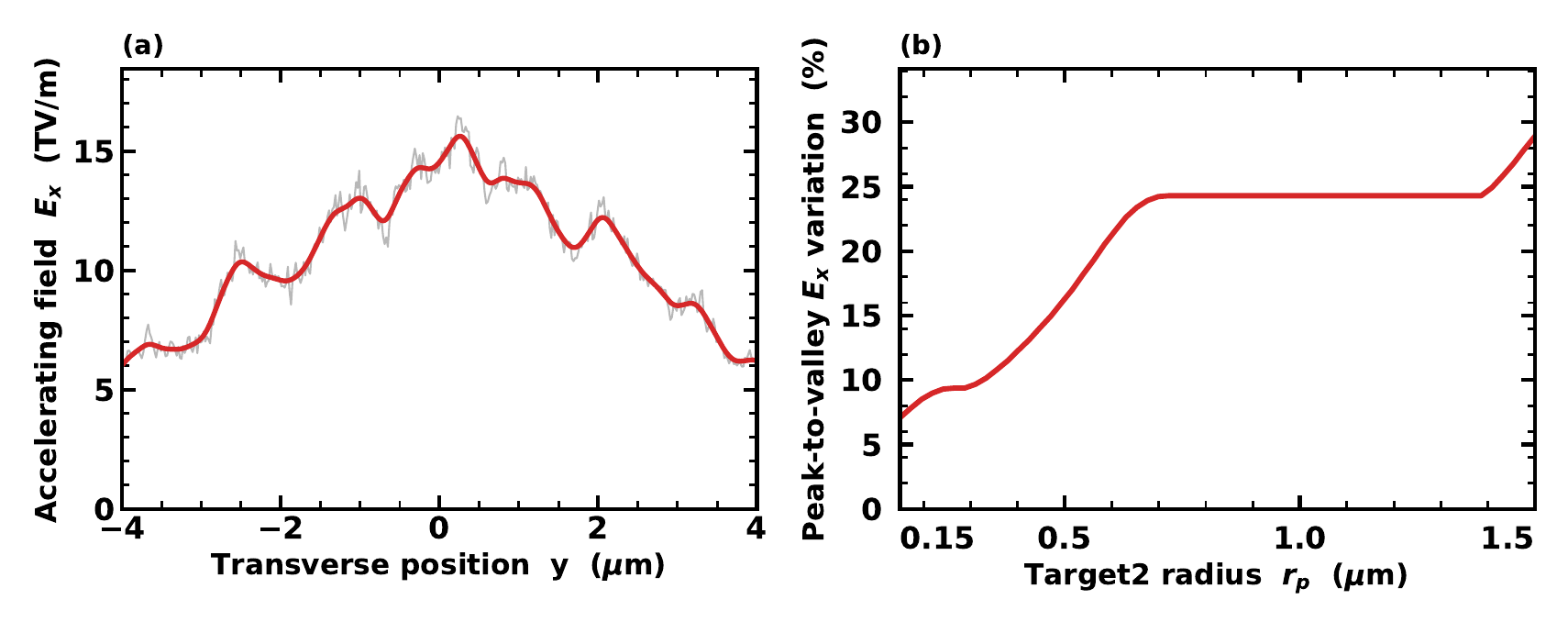}
  \caption{Source-size dependence of the rear accelerating field for evaluation 50. (a) Transverse profile of the longitudinal field $E_x$ across the proton layer at the sheath-formation time ($t\approx100$ fs); (b) Peak-to-valley variation of $E_x$ across a source of radius $r$, increasing from about 7\% at $r_p=0.15\,\mu{\rm m}$ to about 30\% at $r_p=1.5\,\mu{\rm m}$.}
  \label{fig:mechanism}
\end{figure*}

The dependence on the source radius can be read directly from the transverse profile of the rear accelerating field. For evaluation 50 at the sheath-formation time ($t\approx100$ fs), the longitudinal field $E_x$ sampled across the proton layer is plotted against transverse position in \cref{fig:mechanism}(a). The field is nearly flat over a central region of order $1\,\mu{\rm m}$ across, where it stays within roughly ten per cent of its on-axis value of $14.6$ TV/m, and it falls towards the wings as the hot-electron density decreases away from the laser axis. A source of radius $r_p=0.15\,\mu{\rm m}$ is confined to this central region, so its protons sample an almost uniform field: the peak-to-valley variation of $E_x$ across $|y|\le r_p$ is about $7\%$. Extending the source to $r_p=1.5\,\mu{\rm m}$ reaches into the curved part of the profile, where the field has dropped to about $80\%$ of its central value, and the sampled variation rises to about $30\%$. Because the final proton energy is set by the time-integrated field at each transverse position, a larger sampled field variation corresponds to a wider spread of final energies.

The nondominated solutions in this scan lie at the lower bounds $r_p=0.15\,\mu{\rm m}$ and $L_2=30$ nm, a hydrogen layer 300 nm in diameter and 30 nm thick. This follows from the mechanism of \cref{fig:mechanism}, in which only a source this small fits within the central plateau of the rear sheath field and samples a nearly uniform accelerating field, whereas a larger source reaches into the falling wings and broadens the spectrum. 

\begin{figure*}[!ht]
  \centering
  \includegraphics[width=\textwidth]{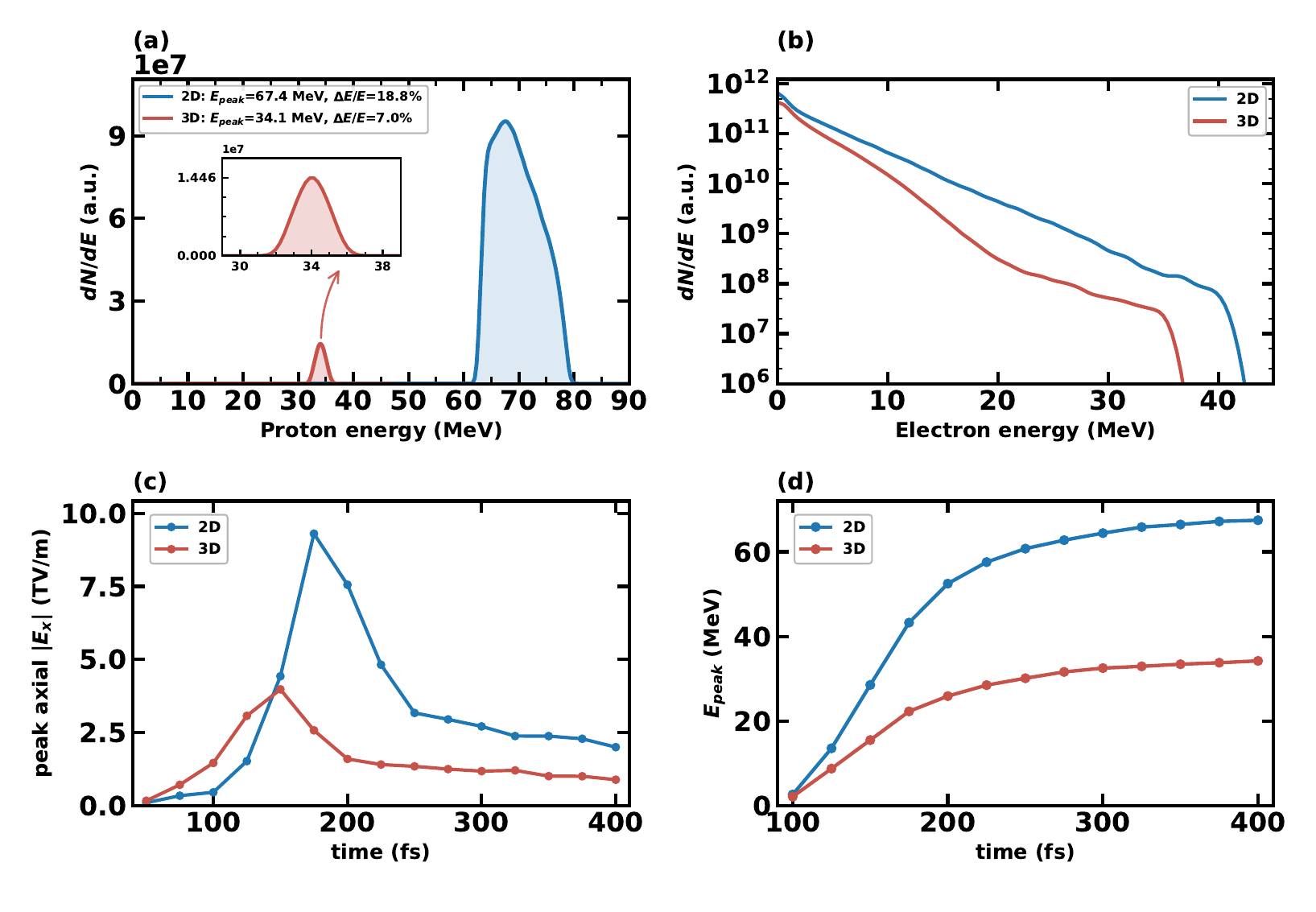}
  \caption{3D simulation of evaluation 50 and acceleration-window mechanism. (a)Proton spectra from the 2D simulation and the corresponding 3D simulation. (b) Rear-side electron spectra at the matched sheath-formation time. (c) Time evolution of the peak axial rear-sheath field $|E_x|$. (d) Time evolution of the proton peak energy. The 3D sheath reaches a lower peak and decays earlier than the 2D sheath, and the 3D proton peak energy saturates earlier.}
  \label{fig:validation}
\end{figure*}

To provide a three-dimensional cross-check, evaluation 50 was repeated and analysed in the same way. The two cases are compared in \cref{fig:validation}. At 400 fs, the 2D simulation gives $E_{\rm peak}=67.4$ MeV, $E_{\rm max}=78.8$ MeV and $\Delta E/E=18.8\%$. The 3D simulation gives $E_{\rm peak}=34.1$ MeV, $E_{\rm max}=36.3$ MeV and $\Delta E/E=7.0\%$. For the 3D spectrum, the half-maximum crossings are at 32.84 and 35.23 MeV, giving ${\rm FWHM}=2.40$ MeV. The bandwidth remains close to this value from 200 to 400 fs, with 6.76\%, 6.78\% and 7.03\% at 200, 300 and 400 fs. At $E_{\rm peak}=34$ MeV the protons have a range of about 1 cm in water, and the quasi-monoenergetic peak gives a well-defined deposition depth. Combined with the high instantaneous flux of a laser-driven source, this suits radiobiology and FLASH-regime irradiation of cell and thin-tissue samples~\cite{zeil2013dose,lv2022flash,kroll2022tumour}. The same energy lies within the range of compact cyclotrons used for medical-radioisotope production, where a narrow energy spread can be matched to a reaction's excitation function to improve yield and purity~\cite{spencer2001laser}.

The peak energy is reduced by a factor 1.97 from 2D to 3D, and the maximum energy by a factor 2.17. The latter is close to the factor of about 2.2 reported in previous 2D optimisation studies of laser-driven proton acceleration when comparing maximum energies to 3D simulations and experiments~\cite{dolier2022multi}. It is also consistent with reports that 2D PIC simulations can overestimate TNSA acceleration because transverse electron spreading and sheath dilution are underrepresented~\cite{stark2017effects,sinigardi2018tnsa}. In two dimensions the hot electrons spread along a single transverse direction, so the rear electron density and the sheath field fall off more slowly than in three dimensions, where the same population expands over a two-dimensional area behind the target. The stronger geometric dilution in 3D lowers the peak sheath field and shortens the interval over which protons are accelerated.

The in-window charge can be compared once the 2D line density is assigned an out-of-plane extent. Taking that extent to be the transverse size of the 3D proton source, $2r_p=0.3\,\mu{\rm m}$, the $27.2$ pC/$\mu{\rm m}$ of evaluation 50 corresponds to $8.2$ pC, against $5.7$ pC obtained directly from the 3D run in the same $\pm10\%$ window. The two agree to within a factor of about 1.4, a weaker reduction than the factor of two in peak energy, with the 2D value the larger in both cases. Both differences are in the direction expected from the stronger transverse expansion of the hot electrons in three dimensions~\cite{stark2017effects,sinigardi2018tnsa}.

These results can be placed against earlier Bayesian-optimisation studies of laser-driven ion sources. Closed-loop experiments have used the method to raise the maximum proton energy~\cite{loughran2023automated}, and numerical campaigns have optimised single energy figures of merit while estimating the move to three dimensions from a fixed literature factor~\cite{dolier2022multi}. Here the bandwidth and the in-window charge are carried through the optimisation as separate objectives, and one parameter set is taken into three dimensions directly, which returns a 2D-to-3D energy factor close to that estimate. The multi-objective, Gaussian-process search itself follows the framework developed for laser-wakefield electron beams~\cite{irshad2023multi}, applied here to the shape of the proton energy spectrum.

The narrower spectrum in 3D is linked to the time evolution of the sheath. As shown in \cref{fig:validation}(c), the peak axial rear-sheath field in 3D is lower and decays earlier than in 2D, and panel (d) shows the proton peak energy saturating correspondingly earlier, whereas the 2D peak continues to gain energy for a longer time. The shorter acceleration window reduces the time over which protons at different positions accumulate different energies, consistent with the smaller late-time bandwidth in the 3D spectrum.

At a comparable pulse energy, broadband TNSA sources deliver of order a nanocoulomb of total proton charge~\cite{daido2012review,macchi2013ion}, but this is spread over a wide, low-energy-dominated spectrum, and downstream energy selection narrows the spectrum only by discarding most of the beam~\cite{brack2020spectral}. Here the charge is concentrated into the peak at the source, the small rear layer sampling the uniform central region of the sheath, so the in-window charge of about 5 pC at $\Delta E/E=7\%$ is obtained without a separate selector. Because this in-window charge rises with the rear-layer density while the peak energy is set by the laser, the same multi-objective search carried out at higher pulse energy and over wider parameter ranges is a possible route to higher-charge quasi-monoenergetic beams in the future.

\section{Conclusion}\label{sec:conclusion}

We have presented a multi-objective Bayesian optimisation of a double-layer target for quasi-monoenergetic TNSA proton acceleration. The optimisation used 80 2D EPOCH simulations and treated the peak energy, spectral purity and in-window charge as separate objectives. 

The Pareto-optimal solutions are concentrated at the bounds $a_0=30$, $\tau=45$ fs, $L_1=0.3\,\mu{\rm m}$, $L_2=30$ nm and $r_p=0.15\,\mu{\rm m}$. The small radius keeps the proton source inside the flat central part of the transverse sheath field, where the accelerating field varies by only a few per cent across the source, against about $30\%$ for a ten-times-larger radius. Within this high-energy subset, the hydrogen-layer density $N_2$ is the main remaining control parameter. Increasing $N_2$ raises the charge in the peak window while increasing the bandwidth, with the 2D peak energy remaining near 64--71 MeV. In the present data set, $N_2$ sets most of the charge--bandwidth variation along this branch.

A 3D simulation of evaluation 50 gives $E_{\rm peak}=34.1$ MeV and $\Delta E/E=7.0\%$ at 400 fs. The lower peak energy is consistent with previous reports that 2D TNSA simulations can overestimate acceleration. The narrower 3D spectrum is associated with earlier decay of the rear sheath and a shorter acceleration window. The quasi-monoenergetic feature is retained in 3D for this parameter set. Further work may sample the same branch and extend the optimisation to higher pulse energy to raise the in-window charge of the quasi-monoenergetic beam.

\ack{This work is supported by the National Natural Science Foundation of China (NSFC) under Grants No. 12375240 and No. 12535015, and by the Program of China Scholarship Council (Grant CSC202506040219).}

\section*{Conflict of Interest}
The authors declare no conflicts of interest.

\data{The data that support the findings of this study are available from the corresponding author upon reasonable request.}

\roles{\textbf{Cheng-Qi Zhang:} Conceptualization (lead); Methodology (lead); Software (lead); Investigation (lead); Formal analysis (lead); Data curation (lead); Writing original draft (lead). \\
\textbf{Yang He:} Methodology (supporting); Writing review \& editing (supporting). \\
\textbf{Mamat Ali Bake:} Formal analysis (supporting); Writing review \& editing (supporting). \\
\textbf{Bai-Song Xie:} Supervision (lead); Funding acquisition (lead); Writing review \& editing (lead).}

\bibliographystyle{unsrt}
\bibliography{ref}

\end{document}